\documentclass{aa}

\usepackage{natbib}
\usepackage{amsmath}
\usepackage{amsfonts}
\usepackage{amssymb}
\usepackage{graphicx}
\usepackage[titletoc]{appendix}
\usepackage{color}
\usepackage{hyperref}
\usepackage{cleveref}
\usepackage[rightcaption]{sidecap}
\usepackage{subfigure}
\usepackage{comment}
\usepackage{soul}
\usepackage{cancel}
\usepackage{dcolumn}

\usepackage{array}
\usepackage{ctable}
\usepackage{multirow}
\usepackage{siunitx}
\usepackage{longtable}
\usepackage{tabularx}
\usepackage{booktabs}
\usepackage{float}

\graphicspath{{Graphics/}}

\def\be{\begin{equation}}
\def\ee{\end{equation}}
\def\bea{\begin{eqnarray}}
\def\eea{\end{eqnarray}}

\def\apjl{ApJ Lett.}
\def\apjs{ApJ Suppl. Ser.}

\def\aap{A\&A}

\definecolor{vividviolet}{rgb}{0.62, 0.0, 1.0}
\definecolor{amaranth}{rgb}{0.9, 0.17, 0.31}
\definecolor{palatinateblue}{rgb}{0.15, 0.23, 0.89}
\definecolor{brightpink}{rgb}{1.0, 0.0, 0.5}
\definecolor{cornflowerblue}{rgb}{0.39, 0.58, 0.93}
\definecolor{deepcarminepink}{rgb}{0.94, 0.19, 0.22}
\definecolor{radicalred}{rgb}{1.0, 0.21, 0.37}

\hypersetup{ linktoc=all,
    colorlinks, linkcolor={palatinateblue},
    citecolor={brightpink}, urlcolor={amaranth}
}

\DeclareUnicodeCharacter{2212}{\ensuremath{-}}

\begin{document}

\title{The afterglow of gamma-ray burst -- supernova connections}

\author{Marco Muccino\inst{1,2,3,4}
\and
Rahim Moradi\inst{5}
\and 
Yu Wang\inst{4,6,7,8}}

\institute{Universit\`a di Camerino, Divisione di Fisica, Via Madonna delle carceri 9, 62032 Camerino, Italy.
\and
Al-Farabi Kazakh National University, Al-Farabi av. 71, 050040 Almaty, Kazakhstan.
\and
INAF - Catania Astrophysical Observatory, Via S.Sofia 79, 95123 Catania, Italy.
\and
ICRANet, Piazza della Repubblica 10,  65122 Pescara, Italy.
\and 
State Key Laboratory of Particle Astrophysics, Institute of High Energy Physics, Chinese Academy of Sciences, Beijing 100049, China.
\and
ICRA, Dipartimento di Fisica, Sapienza Universit\'a di Roma, Piazzale Aldo Moro 5, I-00185 Roma, Italy.
\and
ICRANet-AI, Brickell Avenue 701, Miami, FL 33131, USA.
\and 
INAF – Osservatorio Astronomico d’Abruzzo, Via M. Maggini snc, I-64100, Teramo, Italy.\\
\email{marco.muccino@unicam.it, rahim.moradi@ihep.ac.cn, yu.wang@icranet.org}}

\date{}

\abstract
{The X-ray afterglow of several long gamma-ray bursts (LGRBs) associated with broad-lined type Ib/c supernovae (SNe) exhibits a standard non-thermal afterglow, commonly attributed to synchrotron emission from a relativistic jet, and an evolving thermal component whose physical origin is still debated.}
{We investigate whether the thermal and non-thermal components observed in LGRB--SN systems can be described within a common analytical framework and whether their temporal evolution can provide insight into the physical connection between relativistic jets and SN ejecta.}
{We combine a phenomenological description of the synchrotron emission of the non-thermal energy reservoir associated with the relativistic jet with a diffusion model describing the thermal evolution of the jet-affected portion of the SN ejecta. The resulting coupled system is solved analytically under suitable approximations, leading to closed-form expressions for both non-thermal and thermal luminosities. We apply the model to the following systems: GRB~060218A/SN~2006aj and GRB~171205A/SN~2017iuk.}
{The proposed formalism reproduces the main features of both thermal and non-thermal light curves. In both systems, the thermal luminosity follows a temporal evolution similar to that of the non-thermal component up to the plateau phase. 
The inferred thermal energy stored in the jet-affected ejecta is found to be a fraction of the energy coupled to the observed non-thermal emission. The characteristic timescales obtained from the fits suggest a direct link between the evolution of the relativistic outflow and the thermal response of the expanding ejecta.}
{Our results support a scenario in which the thermal and non-thermal emissions trace different manifestations of the same engine-driven phenomenon. Although simplified, the proposed framework provides a unified analytical description of LGRB afterglows and their thermal counterparts and offers a useful tool for investigating the physical connection between relativistic jets and SNe Ib/c.}


\titlerunning{The afterglow of LGRBs--SNe}

\authorrunning{M.~Muccino et al.}

\maketitle

\section{Introduction}\label{sec:intro}

It is generally accepted that in SN explosions a shock wave propagates from the collapsing core of the star through the stellar envelope. As the shock breaks out, it leaves behind a radiation-dominated gas whose subsequent evolution is governed by the diffusion of photons through the expanding ejecta \citep[see, e.g.,][and references therein]{Arnett1982}. Except for the initial phase of acceleration due to the work of the gas, the expansion rapidly approaches a homologous regime and can be approximated by a constant velocity field $v$.
The evolution of SN light curves has been extensively studied both numerically and analytically. Type II SNe can be described through numerical simulations \citep{FalkArnett1973,FalkArnett1977,Chevalier1976} and analytical calculations \citep{Arnett1980} of explosions occurring in extended stellar envelopes. These studies showed that only a small fraction of the explosion energy is emitted as radiation, while most of it is converted into kinetic energy of the ejecta. A further development of this framework was presented by \citet{Arnett1982}, who included the effects of radioactive heating from the decay of $^{56}$Ni and $^{56}$Co.

The association between LGRBs and broad-line type Ib/c SNe (SNe Ib/c) is now well established observationally \citep{Hjorth,Dellavalle}. These systems are characterized by stripped-envelope progenitors, high explosion energies, and ejecta velocities reaching $\approx0.3c$. Their observational properties indicate the presence of an additional energy reservoir, commonly associated with a relativistic outflow powered by a central engine. The nearby events GRB~980425/SN~1998bw, GRB~060218/SN~2006aj, and GRB~171205A/SN~2017iuk provide some of the best-studied examples of this class.

In addition to the non-thermal X-ray afterglow commonly interpreted as synchrotron emission from the relativistic jet \citep{Sari1998b,Zhang2006}, several LGRBs exhibit an evolving thermal component during the early phases of their X-ray emission \citep{2012MNRAS.427.2950S}. The origin of this component is still debated. Proposed interpretations include shock breakout emission \citep{Campana:2006qe}, radiation from a cocoon surrounding the jet, and energy dissipation resulting from the interaction between the jet and the SN ejecta. Regardless of its physical origin, the thermal component provides a direct probe of the connection between the relativistic outflow and the expanding stellar material.

Most studies have treated the non-thermal and thermal emissions separately. In this work, we investigate the possibility that both components originate from a common physical framework. 
We combine a phenomenological description of the fraction of jet energy coupled to the observed non-thermal emission with a diffusion model describing the thermal evolution of the portion of SN ejecta affected by the jet propagation. The resulting formalism provides a unified analytical description of the X-ray afterglow and its thermal counterpart. We apply the model to the prototypical systems GRB~060218/SN~2006aj and GRB~171205A/SN~2017iuk and discuss the implications for the physical connection between LGRBs and SNe Ib/c.

The paper is organized as follows.
Sec.~\ref{sec:2} introduces the theoretical setup.
Sec.~\ref{sec:3} applies our physical framework to the systems GRB~060218/SN~2006aj and GRB~171205A/SN~2017iuk.
Sec.~\ref{sec:4} draws conclusions and perspectives of this work.

\section{Theoretical setup}\label{sec:2}

The spectral energy distribution (SED) of the X-ray afterglow emission is usually well fitted by a power-law model over the entire observational time. 
However, in some sources an extra thermal component is required to better reproduce their SED, especially at early times, e.g., $t\lesssim10^3$ s.
Therefore, a complete description of the X-ray afterglow should take into account both non-thermal (n) and thermal (t) spectral components. 

The non-thermal part originates from the LGRB jet, representing the bulk of the X-ray afterglow, is explained by the synchrotron emission \citep[see][for details]{Sari1998b}, whereas the thermal part describes the SN emission \citep{Arnett1980}, which is powered by energy deposited by the jet/ejecta interaction and is assumed to trace the time dependence of the central-engine injection.
Non-thermal and thermal parts are described, respectively:
\begin{subequations}
\label{thermo}
\begin{align}
\label{thermo1b} 
& \dot{E}_{n} = - \frac{E_{n}}{\tau_c} + \epsilon_n\,,\\
\label{thermo1a} 
& \dot{E}_{t} + P_{t}\dot{V}_t = - \frac{\partial L_{t}}{\partial m} + \epsilon_t\,.
\end{align}
\end{subequations}
where we define internal energies $E_x$, luminosities $L_x$, and source terms $\epsilon_x$ of both non-thermal and thermal components $x=\{n,t\}$.\footnote{Note that Eq.~\eqref{thermo1a} is written in Lagrangian mass coordinates, but later in Sec.~\ref{sec:2.2} the correct physical units are restored.}
The quantity $E_n$ should be interpreted as the fraction of the jet energy effectively coupled to the observed non-thermal radiation, rather than the total kinetic energy of the outflow, whereas $E_t$ represents the internal energy of the portion of SN ejecta directly affected by the jet energy injection, rather than the total internal energy of the SN ejecta.
The pressure $P_t$ and the mass element $m$ refers to the SN ejecta, whereas the cooling time $\tau_c$ to the non-thermal part. 
The radii $R_x$ and the volumes $V_x$ of the above components are different because typically SN ejecta expands at mildly relativistic velocities, whereas LGRBs are ultra-relativistic sources.

\subsection{Evolution of the non-thermal component}\label{sec:2.1}

In the standard forward-shock model, the X-ray afterglow originates in a relativistic shock between the LGRB jet and the cold external medium, at distances $\sim10^{16}$--$10^{17}$~cm \citep[see e.g.][]{Sari1998b,Piran2005,Meszaros2006}. The synchrotron emission arises from the electrons accelerated by the magnetic field $B$ associated with the turbulent plasma and generated in the relativistic shock \citep[see, e.g.,][]{Sari1998b,Granotetal2000}.

Hereby, we adopt the synchrotron mechanism, but with some fundamental differences from the standard model.
\begin{itemize}
\item[1.]{The LGRB jet energizes and accelerates the SN ejecta.}
\item[2.]{The jet propagates through the SN ejecta, 
breaks out from the stellar surface or from a dense circumstellar environment, and a relativistic shock occurs.}
\item[3.]{The synchrotron emission arises as the shock propagates through the external medium from the progenitor star.}
\end{itemize}

As long as the swept-up mass is negligible, the jet expansion velocity is homologous ($\beta_j={\rm const}$), so that 
\begin{equation}
\label{homolog} 
R_n(t)=D_0+\beta_j c t = D_0\left(1+\frac{t}{\tau_j}\right)\,,
\end{equation}
where $t$ is the source rest-frame time, and $\tau_j=D_0/(\beta_j c)$ the expansion time scale.

Next, we assume electrons accelerated to a power-law distribution of Lorentz factors $\propto\gamma^{-p}$, with $\gamma\in[\gamma_1,\gamma_2]$ and index $p$. Thus, the electron number density is
\begin{equation}
\label{eldistr} 
n(\gamma,t)d\gamma=n_0\left(1+\frac{t}{\tau_j}\right)^{-m}\gamma^{-p}d\gamma\ ,
\end{equation}
where $n_0$ is the initial value and the index $m=0,2,3$ indicates constant, wind and matter distribution, respectively.

We consider a magnetic field decaying from an initial value $B_0$ as the expansion of the ejecta goes on
\begin{equation}
\label{magfield} 
B(t)=B_0\left(1+\frac{t}{\tau_j}\right)^{-b}\ .
\end{equation}

The synchrotron power per unit volume and frequency $\nu$ is given by the  convolution \citep{RybickiLightman}
\begin{equation}
\label{synchrotron} 
F_\nu(\nu,t)= \int_{\gamma_1}^{\gamma_2}{\mathcal{F}(\nu,\gamma,t)n(\gamma,t)} d\gamma\,,
\end{equation}
in which the single-electron synchrotron emissivity is
\begin{equation}
\label{synchrotronmel} 
\mathcal{F}(\nu,\gamma,t)=\frac{\sqrt{3}q^3B(t)\Gamma_j\sin\varphi}{m_ec^2}\frac{\nu}{\nu_c}\int_{\frac{\nu}{\nu_c}}^{+\infty}{K_{5/3}(x)dx}\,,
\end{equation}
and the critical synchrotron frequency is given by 
\begin{equation}
\label{critfreq} 
\nu_c(\gamma,t)=\frac{3qB(t)\gamma^2}{4\pi m_ec}\Gamma_j\sin\varphi\,,
\end{equation}
where $\Gamma_j=(1-\beta_j^2)^{-1/2}$ is the bulk Lorentz factor,
$K_{5/3}(x)$ is a second kind modified Bessel function, and $q$, $m_e$ and $\varphi$ are, respectively, the electron charge, mass and \textit{pitch angle} between the field and the electron velocity. 

The synchrotron luminosity can be obtained by integrating Eq.~\eqref{synchrotron} over the rest-frame X-ray frequencies, averaging over the $\varphi$ angle distribution, and multiplying by the volume of the system $V_n(t)=4\pi D_0^3/3(1+t/\tau_j)^3$. Detailed calculations in App.~\ref{appe:1} provide the source term of Eq.~\eqref{thermo1b}
\begin{equation}
\label{muccillingale} 
\epsilon_n(t)=\epsilon_n(0)\left(1+\frac{t}{\tau_j}\right)^\alpha\,,
\end{equation}
where $\epsilon_n(0)$ and $\alpha<0$ are constants.

\begin{figure*}
\centering
\includegraphics[width=0.49\hsize,clip]{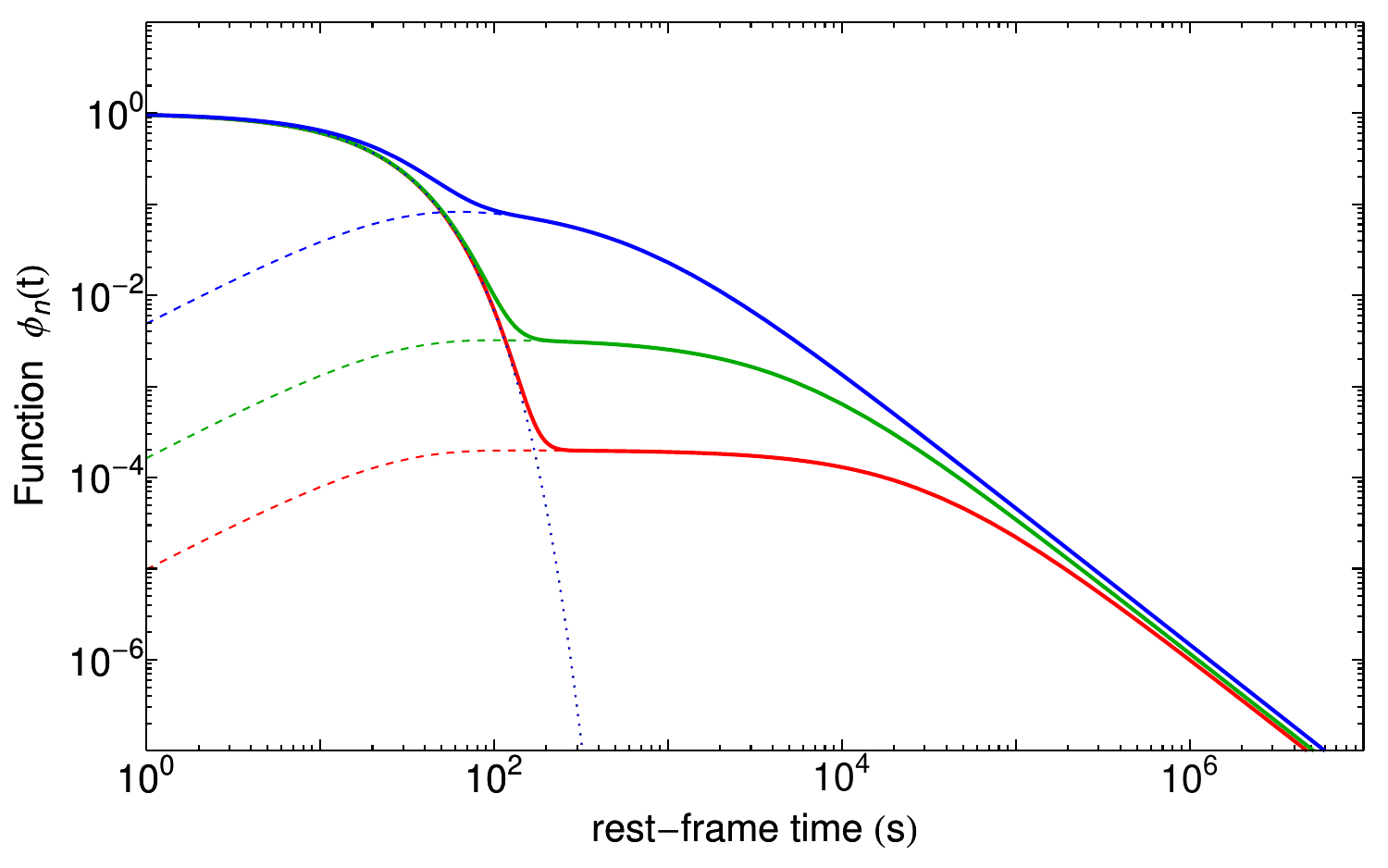}\hfill
\includegraphics[width=0.49\hsize,clip]{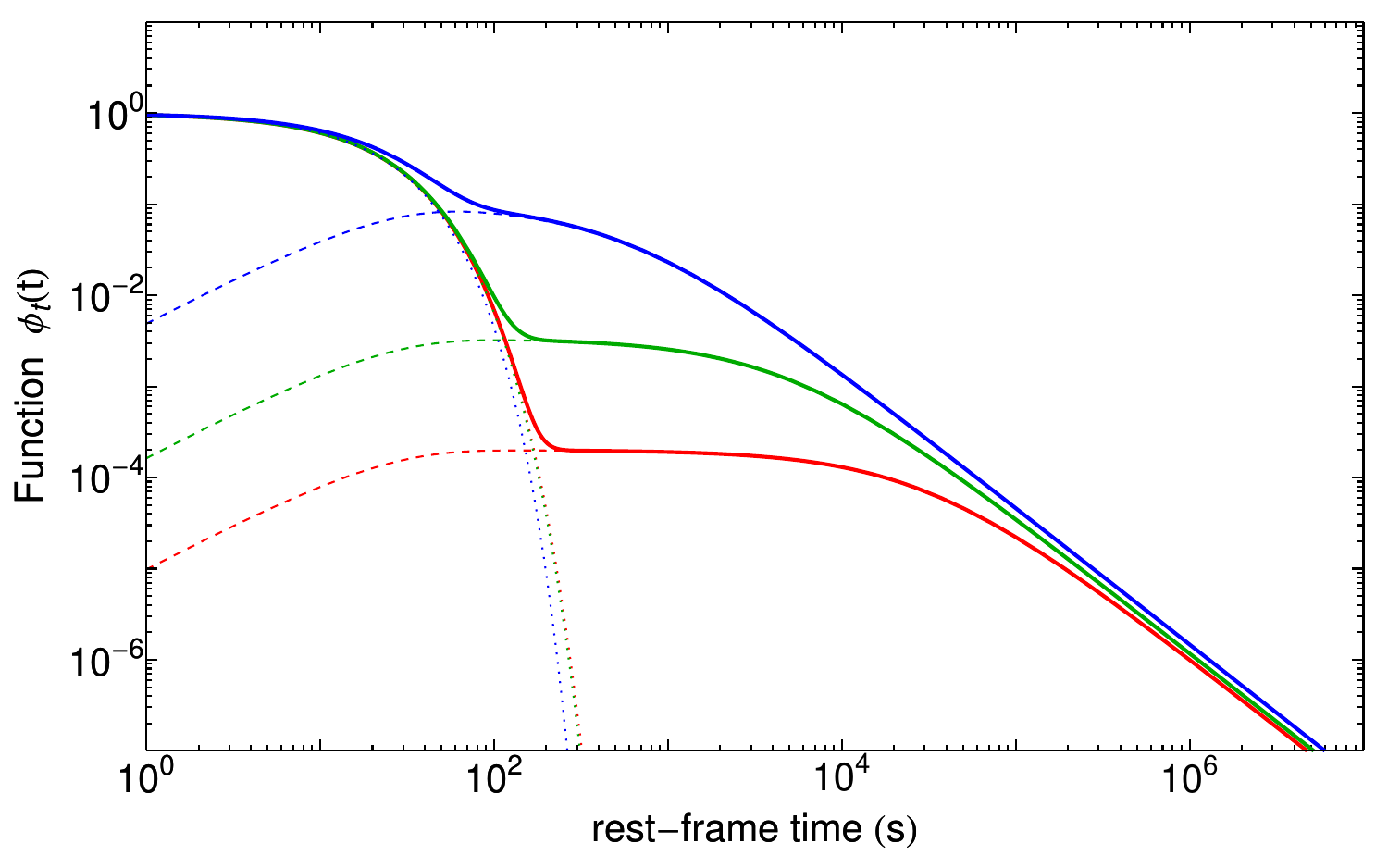}
\caption{{\it Left panel}: plot of $\phi_n(t)$ obtained by fixing $\tau_c=20$~s and $\alpha=-1.5$, and setting: $(\tau_j,\,\tau_n)=(600\,{\rm s},\,200\,{\rm s})$ (blue curve), $(5000\,{\rm s},\,6000\,{\rm s})$ (green curve), and $(30000\,{\rm s},\,10^5\,{\rm s})$ (red curve). {\it Right panel}: plot of $\phi_t(t)$ obtained by fixing $\tau_0=20$~s and $\alpha=-1.5$, and setting $(\tau_h,\,\tau_t)$ with the same values of $(\tau_j,\,\tau_n)$ and the same color and symbol code.}
\label{fig:1}
\end{figure*}

Finally, we consider the jet to be homogeneous, so that the non-thermal internal energy is only time-dependent,
\begin{equation}
\label{ntpar} 
E_n(t)=E_n(0)\phi_n(t)\,,
\end{equation}
and plugging it and Eq.~\eqref{muccillingale} into Eq.~\eqref{thermo1b} and we obtain
\begin{equation}
\label{muccillingale2} 
\dot\phi_n(t) + \frac{\phi_n(t)}{\tau_c} = \frac{1}{\tau_n}\left(1+\frac{t}{\tau_j}\right)^\alpha\,,
\end{equation}
where we defined the source time scale $\tau_n=E_n(0)/\epsilon_n(0)$. With no loss in generality, we set the initial condition $\phi_n(0)=1$. The solution of Eq.~\eqref{muccillingale2} is well approximated within $\approx4\%$ of accuracy (see details in App.~\ref{appe:2}) by
\begin{equation}
\label{muccillingale3} 
\phi_n(t) = e^{-t/\tau_c} + \frac{\tau_c}{\tau_n}\left(1-e^{-t/\tau_c}\right)\left(1+\frac{t}{\tau_j}\right)^\alpha\,.
\end{equation}
Fig.~\ref{fig:1} (left panel) portrays Eq.~\eqref{muccillingale3}, where the first term (see dotted curves) is the steep decay observed in X-ray afterglows of LGRBs \citep{Zhang2006} and the second one (see dashed curves) reproduces the plateau \citep{RuffiniMuccino2014} and the late power-law decay \citep{Pisani2013}.
 
Finally, from Eqs.~\eqref{ntpar} and \eqref{muccillingale3} we can compute the non-thermal luminosity as $E_n(t)/\tau_c$, leading to
\begin{equation}
\label{ntlum} 
L_n(t)= \frac{E_n(0)}{\tau_c}\phi_n(t)\,.
\end{equation}
The second term is the generalization of the function introduced in \citet{RuffiniMuccino2014}, used to derive the Combo correlation for LGRBs \citep{Izzo2015,2021ApJ...908..181M,Muccino:2026gvt}.
See also the derivation in \citet{2017MNRAS.468..570M}.

\subsection{Evolution of the thermal component}\label{sec:2.2}

For the thermal component in Eq.~\eqref{thermo1a}, the internal energy and the pressure are related by $E_{t}/V_t=3P=aT^4$, where $a$ is the radiation constant, and the specific volume $V_t=1/\rho$ depends upon the SN ejecta mass density $\rho$.
We define the mass element as $dm=4\pi r^2\rho dr$, where $r$ is the radial coordinate.
Then, we introduce the following set of approximations \citep{Arnett1980,Arnett1982}:
\begin{itemize}
\item[-] the diffusion approximation for the luminosity  
\begin{equation}
L_t=- 4\pi r^2\left(\frac{ac}{3\rho\kappa}\right)\frac{\partial T^4}{\partial r}\ ;
\label{lumdiff}
\end{equation}
\item[-] the constant Thomson opacity, i.e. $\kappa\equiv\kappa_0$;
\item[-] the homologous expansion of the jet-affected ejecta region (at v={\rm const}), leading to 
\begin{equation}
R_t(t)=R_0\left(1+\frac{t}{\tau_h}\right)\quad,\quad\frac{\dot V_t(t)}{V_t(t)}=3\frac{\dot R_t(t)}{R_t(t)}\,,
\label{homexp}
\end{equation}
where $\tau_h=R_0/v$ is the expansion timescale;
\item[-] the adiabatic approximation $T\propto R_t^{-1}$, i.e.,
\begin{equation}
T(x,t)^4 = T(0,0)^4\psi(x)\phi_t(t)\left[\frac{R_0}{R_t(t)}\right]^4\ .
\label{temp}
\end{equation}
where $\phi_t(t)$ describes deviations from the adiabaticity and $\psi(x)$ includes the spatial dependence parametrized by the dimensionless radial coordinate $x=r/R_t(t)$;
\item[-] the mass density parametrization  
\begin{equation}
\rho(x,t) = \rho(0,0)\eta(x)\left[\frac{R_0}{R_t(t)}\right]^3\ ,
\label{dens}
\end{equation}
where $\eta(x)$ defines the shape of the density distribution;
\item[-] the heat source term due to the LGRB jet 
\begin{equation}
\label{decays}
\epsilon_t(x,t) = \epsilon_t(0,0)\xi(x) 
\left[\frac{R_t(t)}{R_0}\right]^\alpha\,,
\end{equation}
where the spatial distribution $\xi(x)$ may be due to the reprocessing of the synchrotron emission in the SN ejecta.\footnote{Besides the synchrotron emission, additional contributions to the X-ray afterglow may come from newly-born pulsar wind, shock, Fermi mechanism, etc. In the optical bands, at late time, $^{56}$Ni and $^{56}$Co decays shall be considered to describe the optical features of SNe Ib/c.}
\end{itemize}
It is worth noticing that, in the present framework, $R_0$ should be interpreted as the characteristic size of the jet-affected ejecta (cavity, cocoon, reprocessed region, etc.) and not necessarily as the progenitor stellar radius.
Moreover, in the last point it is assumed that the heating of the SN ejecta is ultimately powered by the same engine responsible for the non-thermal afterglow. As the jet propagates through and interacts with the ejecta, a fraction of its dissipated energy is converted into thermal energy. We therefore model the thermal source term with the same functional time dependence as the non-thermal injection term, but with an independent normalization and an independent hydrodynamical timescale. This choice does not imply identical jet and ejecta dynamics; rather, it encodes the assumption that both components trace the same underlying engine activity.

By using Eqs.~\eqref{lumdiff}--\eqref{decays}, Eq.~\eqref{thermo1a} can be written as
\begin{equation}
\dot{\phi}_t(t)+\frac{\phi_t(t)}{\tau_0} \left(1+\frac{t}{\tau_h}\right)=\frac{1}{\tau_t(x)} \left(1+\frac{t}{\tau_h}\right)^{1+\alpha}\,,
\label{partial1}
\end{equation}
in which we have defined the diffusion timescale
\begin{equation}
\tau_0\equiv\frac{3 R_0^2\rho(0,0)\kappa_0}{\theta(x)c}\,, 
\label{tdiff}
\end{equation}
in which we defined
\begin{align}
\label{partial2}
\theta(x)=&\,-\frac{1}{x^2\psi(x)}\frac{\partial}{\partial x}\left[\frac{x^2}{\eta(x)}\frac{\partial \psi(x)}{\partial x}\right]\,,\\
\label{partial3}
\tau_t(x)=&\,\left[\frac{\epsilon_t(0,0)\rho(0,0)}{aT(0,0)^4} \frac{\xi(x)\eta(x)}{\psi(x)}\right]^{-1}\,.
\end{align}
If the specific heat energy $\xi(x)\eta(x)$ proportional to the radiation energy density $\psi(x)$ \citep{Arnett1982}, then $\tau_t$ and $\theta$ are constants and Eqs.~\eqref{partial1} and (\ref{partial2}) are separable.
Therefore, setting again the initial condition $\phi(0)=1$, the solution of Eq.~(\ref{partial1}) is well approximated within $\approx4\%$ of accuracy (see details in App.~\ref{appe:2}) by
\begin{align}
\nonumber
\phi_t(t)=&\,e^{-\frac{t}{\tau_0}\left(1+\frac{t}{2\tau_h}\right)}+\\
\label{timepartsol}
&\,\frac{\tau_0}{\tau_t}\left[1-e^{-\frac{t}{\tau_0}\left(1+\frac{t}{2\tau_h}\right)}\right]\left(1+\frac{t}{\tau_h}\right)^\alpha\,,
\end{align}
and it is displayed in the right panel of Fig.~\ref{fig:1} for the selected values of the parameters $(\tau_j,\,\tau_t,\,\tau_c,\,\alpha$).

As the SN shock breaks out from $R_0$ at $t=0$, the stellar envelope, akin to a radiation-dominated gas, is heated and accelerates.
The energy developed in the shock is equally divided into internal and kinetic energies of the gas, with the internal energy given by
\begin{equation}
E_t(t) = E_t(0)\phi_t(t)\left(1+\frac{t}{\tau_h}\right)^{-1},
\label{intene}
\end{equation}
where $E_t(0)=4\pi R_0^3 aT(0,0)^4I_t$ is the initial internal energy and $I_t=\int^{1}_{0}\psi(x)x^2dx$ a dimensionless factor.

From Eq.~\eqref{lumdiff} the surface luminosity writes as
\begin{equation}
\label{lumsur}
L_t(t) = \frac{4\pi a T(0,0)^4 c R_0}{3\kappa_0\rho(0,0)} \left[-\frac{x^2}{\eta(x)}\frac{\partial \psi(x)}{\partial x}\right]_{x=1}\phi_t(t)\,.
\end{equation}
The quantity in the brackets is evaluated by integrating Eq.~\eqref{partial2} imposing some boundary conditions at the center of the outflow, {\it i.e.}, $\psi(0)=1$ and $(d\psi/dx)_{x=0}=0$, and assuming a constant or slowly varying $\eta(x)$ \cite[see, e.g.,][]{Arnett1982}.
Therefore, with these figures, Eq.~\eqref{lumsur} becomes
\begin{align}
\nonumber
L_t(t) = \frac{E_t(0)}{\tau_0} \phi_t(t)\,.
\end{align}

\section{Application to LGRB--SN connections}\label{sec:3}

We apply the above-described model to reproduce the X-ray afterglow of the LGRBs~060218A and 171205A.

\subsection{LGRB~060218A--SN~2006aj}\label{sec:3.1}
%
\begin{figure*}
\centering
\includegraphics[width=0.49\hsize,clip]{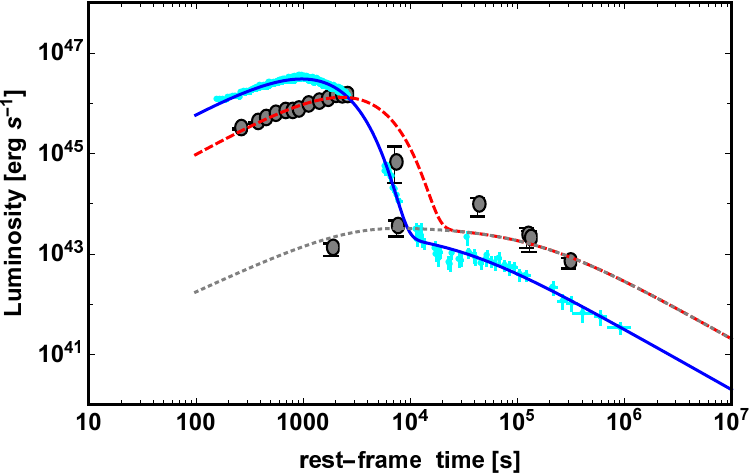}
\includegraphics[width=0.49\hsize,clip]{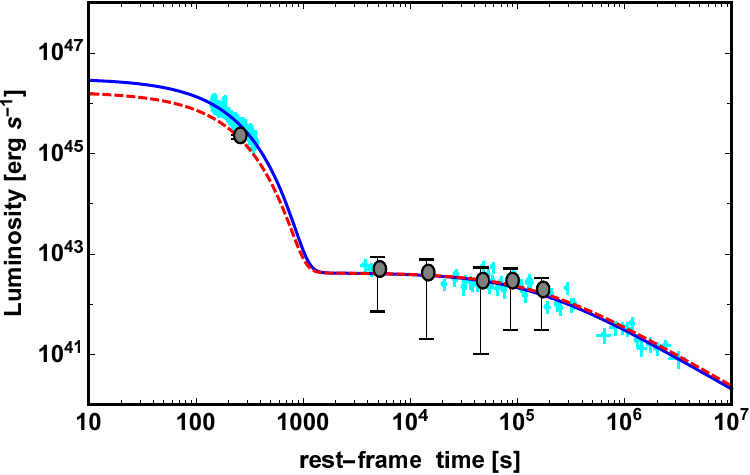}
\caption{Fits of non-thermal (cyan data and blue curve) and thermal (gray data and red-dashed curve) components of LGRBs~060218A and 171205A. For LGRB~060218A some data points on the boosted SN emission have been obtained from early optical data and their behavior is shown with a gray dotted curve. The corresponding parameters are given in Table~\ref{tab:1}.}
\label{fig:2}
\end{figure*}
\begin{table*}
\caption{X-ray afterglow non-thermal and thermal best-fit parameters of LGRBs~060218A and 171205A.}
\centering
\setlength{\tabcolsep}{0.6em}
\renewcommand{\arraystretch}{1.2}
\begin{tabular}{llccccccccc}
\hline\hline
LGRB                            &
$x$                             &
$E_n(0)$                        &
$E_t(0)$                        &
$\tau_c$                        &
$\tau_n$                        &
$\tau_j$                        &
$\tau_0$                        &
$\tau_t$                        &
$\tau_h$                        &
$\alpha$                        \\
                                &
                                &
[$10^{46}$erg]                 &
[$10^{46}$erg]                 &
[s]                             &
[ks]                            &
[ks]                            &
[ks]                            &
[ks]                            &
[ks]                            &
                                \\
\hline   
060218A                         &
n                               & 
$524\pm4$                       &
$-$                             &
$814\pm3$                       & 
$183\pm44$                      &
$24\pm13$                       &
$-$                             &  
$-$                             &
$-$                             &
$-1.21\pm0.16$                  \\
                                &
t                               & 
$-$                             &
$192\pm9$                       &
$-$                             & 
$-$                             &
$-$                             &
$2.0^{+0.1}_{-0.6}$             &  
$56\pm24$                       &
$143\pm114$                     &
$-1.21^\star$                   \\
\hline   
171205A                         &
n                               & 
$370\pm21$                      &
$-$                             &
$121\pm3$                       & 
$889\pm99$                      &
$134\pm43$                      &
$-$                             &  
$-$                             &
$-$                             &
$-1.23\pm0.13$                  \\
                                &
t                               & 
$-$                             &
$200\pm2$                       &
$-$                             & 
$-$                             &
$-$                             &
$0.12^\star$                   &  
$450\pm38$                      &
$150\pm32$                      &
$-1.23^\star$                   \\
\hline
\end{tabular}
\tablefoot{Because of the shortage of data for the thermal components, some parameters have been fixed to the values obtained from the non-thermal component (see starred values).}
\label{tab:1}
\end{table*}

The nearby LGRB~060218 ($z=0.033$), associated to the SN~2006aj, provided the first direct observation of a SN shock breakout associated with a LGRB. The detected thermal component cooling from the X-ray down to the UV/optical band, was interpreted as the radiation released when the SN shock emerged from a compact Wolf-Rayet progenitor and its dense wind. This observation established a direct physical link between the relativistic jet producing the LGRB and the core-collapse explosion producing the SN Ic, confirming that LGRBs and broad-lined SNe Ic are different manifestations of the same stellar death event \citep[see][for more details]{Campana:2006qe}.

The thermal data points shown in Fig.~\ref{fig:2} (left panel) are of two kinds: (a) the measurements with bolometric luminosities $>10^{44}$~erg , reproduced from \citet{Campana:2006qe}; (b) the complementary bolometric luminosities $<10^{44}$~erg, obtained from optical/UV fluxes by applying the method introduced in \citet{Ghisellini:2007ya}, only in the time intervals where all Swift-UVOT filters simultaneously collected data ($t\sim1.8$, $7.4$, $42$, $122$, $129$, and $303$~ks), but not necessarily restricting the fits to the Rayleigh–Jeans part of a black–body spectrum.
The data from the SN bump are not considered because they originate from the emission due to $^{56}$Ni and $^{56}$Co decays that is beyond our model.

We apply Eq.~\eqref{ntlum} for constraining the non-thermal component to the X-ray afterglow data\footnote{\url{https://www.swift.ac.uk/burst_analyser/00191157/}}, duly transformed into the luminosity light curve (LC) in the rest-frame band $0.3$--$10$~keV. 
For the thermal part, the total luminosity is described by the sum of the two terms in Eq.~\eqref{lumsur3}. The early bolometric data points (a) mainly constrain the exponentially declining shock-cooling contribution, whereas the later optical/UV data points (b) are mostly sensitive to the slower component associated with continued energy deposition. This separation should therefore be interpreted as an approximate dominance criterion rather than as a strict decomposition of the observed luminosity.
Accordingly, both terms are included in the fitting function whenever the temporal coverage allows it.

Since both thermal and non-thermal early-time LCs exhibit a peculiar rising before the exponential decline, we include in both fitting functions a term $\propto t^{1.2}$, introduced phenomenologically to reproduce the observed initial rise.

The best-fit curves are shown in Fig.~\ref{fig:2} (left panel) and the numerical results are summarized in Table~\ref{tab:1}. Because of the shortage of late-time thermal data, to perform a meaningful fit we fixed $\alpha$ to the value got from the non-thermal fit, as indicated by the starred quantity in Table~\ref{tab:1}.
Although the non-thermal rest-frame $0.3$--$10$~keV luminosity LC is not bolometric like the thermal one, we can draw the general conclusions listed below.
\begin{itemize}
\item The thermal bolometric LC follows the same behavior of the non-thermal part up to the plateau phase. 
\item Provided that the X-ray luminosity is not bolometric, the jet energy transferred into the non-thermal emission, $E_n(0)$, amounts to $\approx8.5\%$ of the isotropic energy of the burst \citep[$E_{\rm iso}=6.2\times 10^{49}$~erg, see][]{Campana:2006qe}; 
the initial thermal energy $E_t(0)$ stored in the jet-affected SN ejecta is about $\lesssim0.4E_n(0)$.
\item We can constrain the jet bulk Lorentz factor $\Gamma_j$ dynamically by using the ratio between the expansion timescales  $\tau_j/\tau_h$.
A second, more phenomenological estimate can be obtained from the ratio $\tau_c/\tau_0$. Together they provide:
\begin{equation}
\frac{\tau_j}{\tau_h}~{\rm or}~\frac{\tau_c}{\tau_0}= \frac{D_0}{R_0}\frac{v}{\beta_j c}\,.
\end{equation}
If the fitted X-ray emission is produced before significant deceleration, then we can take $\Gamma_j\approx {\rm const}$. Next, assuming the jet to be directed along the line of sight, we approximate $D_0\approx c\Delta t/(1-\beta_j)$, where $\Delta t=2100$~s is the observed duration of LGRB~060218A.
For an order-of-magnitude estimate, we identify $R_0$ with the initial radius $R_0=5.2\times 10^{11}$~cm inferred by \citet{Campana:2006qe}, although in the present framework $R_0$ more generally represents the characteristic size of the jet-affected ejecta.
Finally, we use the SN expansion velocity $v\approx 0.1c$ and the values in Table~\ref{tab:1} we get the following two estimates: $\Gamma_{j,1}=27^{+10}_{-27}$ and $\Gamma_{j,2}=42^{+2}_{-7}$.  
The intersection of these two estimates provides $\Gamma_j=35$--$37$.
\end{itemize}

\subsection{LGRB~171205A--SN~2017iuk}\label{sec:3.2}

LGRB~171205A \citep[$z=0.0368$,][]{2017GCN.22180....1I} is associated to the SN~2017iuk, characterized by extremely high expansion velocities of about $0.3c$ that were interpreted as due to the mildly-relativistic hot cocoon that is generated by an ultra-relativistic jet within the LGRB \citep{2019Natur.565..324I}.

The best-fit parameters, obtained by fitting the rest-frame $0.3$--$10$~keV luminosity LC\footnote{\url{https://www.swift.ac.uk/burst_analyser/00794972/}} with Eq.~\eqref{ntlum} and the bolometric thermal data \citep{2019Natur.565..324I} with Eq.~\eqref{lumsur3}, are summarized in Table~\ref{tab:1}.
The best-fit curves are shown in Fig.~\ref{fig:2} (right panel). Because of the shortage of thermal data, to perform a meaningful fit we fixed $\tau_0$ with $\tau_c$ and used the value of $\alpha$ got from the non-thermal fit, as indicated by the starred quantities in Table~\ref{tab:1}.
The results are summarized as follows.
\begin{itemize}
\item Although with sparse data points, also here the thermal bolometric LC appears to follow the same behavior of the non-thermal one.  
\item The initial energy of the non-thermal component, $E_n(0)$, is also here $\approx15\%$ of the isotropic energy of the burst \citep[$E_{\rm iso}=2.4\times 10^{49}$~erg, see e.g.][]{2019Natur.565..324I}; the initial thermal energy $E_t(0)$ stored in the jet-affected SN ejecta is about $\approx0.5E_n(0)$.
\item The bulk Lorentz factor $\Gamma_j$ here can be estimated only from the following ratio
\begin{equation}
\frac{\tau_j}{\tau_h} = \frac{D_0}{R_0}\frac{v}{\beta_j c}\,.
\end{equation}
We take the observed duration of LGRB~171205A $\Delta t=190$~s, and for an order-of-magnitude estimate, we use the initial expansion velocity $v\approx 0.3c$ and the inferred initial radius $R_0=1.5\times 10^{12}$~cm from SN~2017iuk \citep{2019Natur.565..324I}. However, the resulting constraint $\Gamma_j\approx188^{+33}_{-41}$ is not as precise as in the case of LGRB~060218, where two pairs of temporal scales are compared.
Moreover, it is worth stating that this estimate is strongly model dependent and relies on the identification of $D_0$ with the characteristic longitudinal size of the emitting region.
\end{itemize}

\section{Conclusions}\label{sec:4}

In this work, we have proposed a unified analytical framework to describe the thermal and non-thermal components observed in the X-ray afterglows of LGRBs associated with broad-line SNe Ib/c. 
The model combines a phenomenological description of the synchrotron emission powered by the relativistic jet with a diffusion model describing the thermal evolution of the portion of SN ejecta directly affected by the jet propagation. Under suitable approximations, both components admit simple analytical solutions that can be directly compared with observations.

We applied this formalism to the nearby systems GRB~060218A/SN~2006aj and GRB~171205A/SN~2017iuk. In both cases, the model successfully reproduces the observed temporal evolution of the non-thermal X-ray afterglow together with the accompanying thermal component. Despite the limited number of thermal measurements, the inferred temporal evolution suggests that the thermal luminosity follows the same overall behavior as the non-thermal component up to the plateau phase, supporting the hypothesis that both emissions are ultimately powered by the same central engine.
Within our interpretation, the non-thermal and thermal components trace different energy reservoirs. The non-thermal emission originates from the fraction of the jet energy converted into the observed synchrotron radiation, whereas the thermal emission arises from the internal energy stored in the jet-affected portion of the SN ejecta. For both systems, the inferred thermal energy is only a fraction of the energy coupled to the observed non-thermal emission, consistently supporting a scenario in which only a limited portion of the SN ejecta is directly heated by the relativistic outflow.
The comparison between the characteristic timescales of the two components provides additional physical insight. Although the estimates remain model dependent and rely on simplifying assumptions regarding the geometry and dynamics of the emitting regions, they suggest that the observed thermal evolution retains memory of the jet activity. In particular, the characteristic scale $R_0$ inferred in the present framework should be interpreted as the effective size of the jet-affected thermal-emitting region, rather than the progenitor stellar radius itself.

The present model is intentionally simplified and several aspects deserve further investigation. The heating term adopted for the thermal component has been introduced phenomenologically by assuming that its temporal evolution follows the same engine activity driving the non-thermal emission. A more complete treatment should derive this coupling self-consistently from relativistic hydrodynamical simulations of jet propagation through the stellar ejecta, including cocoon formation, radiative transfer, and possible departures from spherical symmetry. Likewise, the assumption of homologous expansion and constant opacity may become inadequate during the earliest phases of the evolution.

Despite these limitations, the proposed framework provides a simple analytical tool for investigating the physical connection between relativistic jets and SN ejecta. Future applications to a larger sample of nearby LGRB--SN systems, as well as comparisons with numerical simulations and multi-wavelength observations, will allow the robustness of the present interpretation to be assessed and may help clarify the physical origin of the thermal emission accompanying engine-driven SNe.

\begin{acknowledgements}
This work made use of data supplied by the UK \emph{Swift} Science Data Center at the University of Leicester.
\end{acknowledgements}

\bibliographystyle{aa}
\bibliography{bibliography}

\clearpage

\begin{appendix}

\onecolumn
\raggedbottom

\section{The synchrotron luminosity}\label{appe:1}

For a sufficiently wide range of the electron Lorentz factors, we approximate $\gamma_1\rightarrow0$ and $\gamma_2\rightarrow +\infty$ \citep{RybickiLightman}, so that the convolution of Eq.~\eqref{synchrotron} gives
\begin{equation} 
\label{synchrotronapprox}
F_\nu(\nu,t)= \frac{\sqrt{3}\Gamma_j^{s+1} q^3 B_0 n_0 \sin^{s+1}\varphi}{m_ec^2(p+1)}\bar{\Gamma}\left(\frac{3p+19}{12}\right) \bar{\Gamma}\left(\frac{3p-1}{12}\right)
\left[\frac{2\pi m_ec\nu}{3qB_0}\right]^{-s}\left(1+\frac{t}{\tau_j}\right)^{-b(1+s)-m}\,,
\end{equation}
where $\bar{\Gamma}$ is the Gamma function and $s=(p-1)/2$. 
To obtain the observed photon spectrum $N(E)\propto E^{-2}$ (equivalently, $F_\nu\propto\nu^{-1}$), we fix $p=3$. 

Next, we average over all the possible pitch angles through the integration $(4\pi)^{-1}\int_{0}^{\pi}{2\pi\sin^3\varphi d\varphi}=2/3$ and integrate Eq.~\eqref{synchrotronapprox} over the rest-frame $0.3$--$10$ keV energy band (in frequencies: $\nu_1=7.2 \times 10^{16}/(1+z)$ Hz and $\nu_2=2.4 \times 10^{18}/(1+z)$ Hz, where $z$ is the source redshift).

Finally, multiplying by $V_n(t)$ we obtain
\begin{equation}
\label{synchlum} 
\epsilon_n(t)=\frac{4\xi q^4 B_0^2 n_0 D_0^3\Gamma_j^2}{\sqrt{3}m_e^2c^3}\int_{\nu_1}^{\nu_2}{\frac{d\nu}{\nu}}\left(1+\frac{t}{\tau_j}\right)^{3-2b-m}\,,
\end{equation}
where $\xi=(1/4)\bar\Gamma(4/3)\bar\Gamma(2/3)$. 
If we label with $\epsilon_n(0)$ all the constant terms and define the index $\alpha=3-2b-m<0$, we obtain the function displayed in Eq.~\eqref{muccillingale}.

\section{Approximated solutions}\label{appe:2}

The analytic solution of Eq.~\eqref{muccillingale2} is
\begin{equation}
\label{muccillingale_exact} 
\phi_n(t) = e^{-y} + e^x \frac{\tau_c}{\tau_n}  \left[E_{-\alpha}(x+y) - \frac{x^{1+\alpha} E_{-\alpha}(x)}{(x+y)^{1+\alpha}}\right]\,.
\end{equation}
where $E_{-\alpha}(x)$ and $E_{-\alpha}(x+y)$ are Generalized Exponential Integral (GEI) functions and we have defined 
\begin{equation}
\label{muccillingale_exact_def} 
x(t) = -\frac{t+\tau_j}{\tau_c}\quad,\quad y(t) = \frac{t}{\tau_c}\,.
\end{equation}
Next, we apply the property of GEI function
\begin{equation}
\nonumber
E_{-\alpha}(x) = x^{1+\alpha} \bar\Gamma\left(1+\alpha,x\right)\qquad,\qquad E_{-\alpha}(x+y) = (x+y)^{1+\alpha} \bar\Gamma\left(1+\alpha,x+y\right)\,,
\end{equation}
where $\bar\Gamma$ is the Incomplete Gamma (IG) function. Defining $\Delta\bar\Gamma=\bar\Gamma(1+\alpha,x)-\bar\Gamma(1+\alpha,x+y)$, Eq.~\eqref{muccillingale_exact} becomes
\begin{equation}
\label{muccillingale_exact2} 
\phi_n(t) = e^{-y} + \frac{\tau_c}{\tau_n} \frac{e^x\Delta\bar\Gamma}{{(x+y)^{1+\alpha}}}\,.
\end{equation}
Now we resort an expansion of the IG functions holding for $|x|>|y|$ (or equivalently $\tau_j>\tau_c$)
\begin{equation}
\label{delta_gamma}
\Delta\bar\Gamma\approx e^{-x}x^\alpha \sum_{n=0}^{+\infty}\frac{(-\alpha)_n}{(-x)^n}\left[1-e^{-y}e_n(y)\right],
\end{equation}
where $(-\alpha)_n=\bar\Gamma(n-\alpha)/\bar\Gamma(-\alpha)$ is the Pochhammer symbol and $e_n(y)=\sum_{k=0}^{n}y^k/(k!)$. By retaining the lower order ($n=0$) of the above expansion, we get Eq.~\eqref{muccillingale3}.

\begin{figure}
\centering
\includegraphics[width=0.48\hsize,clip]{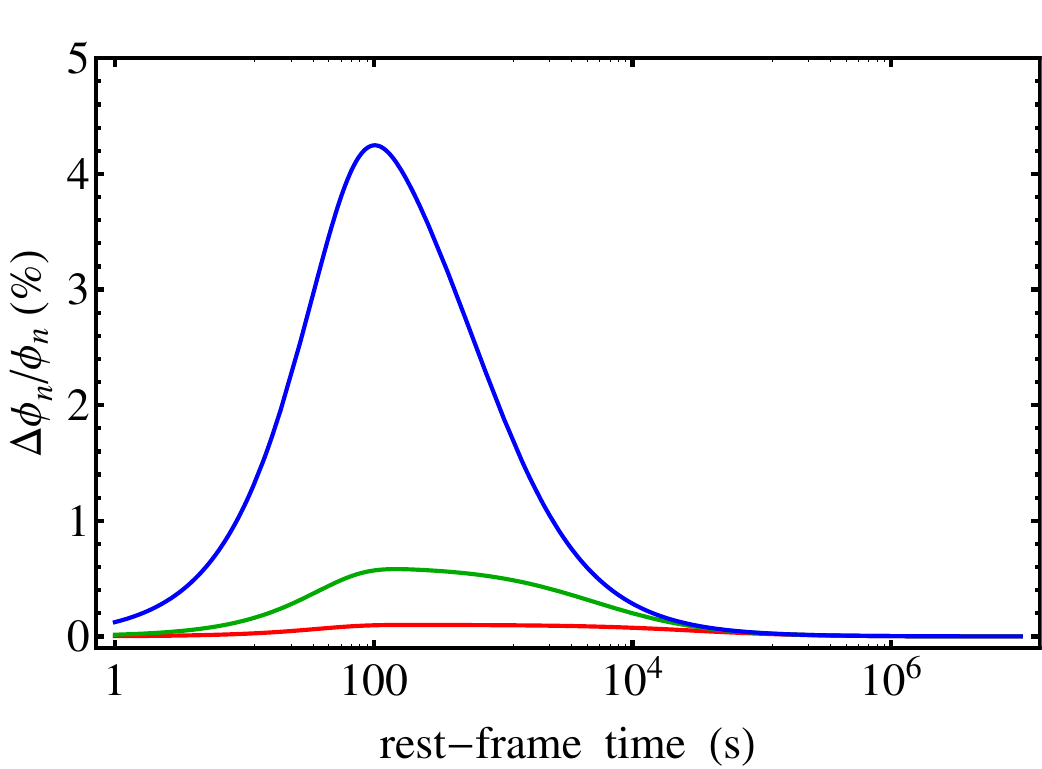}\hfill
\includegraphics[width=0.48\hsize,clip]{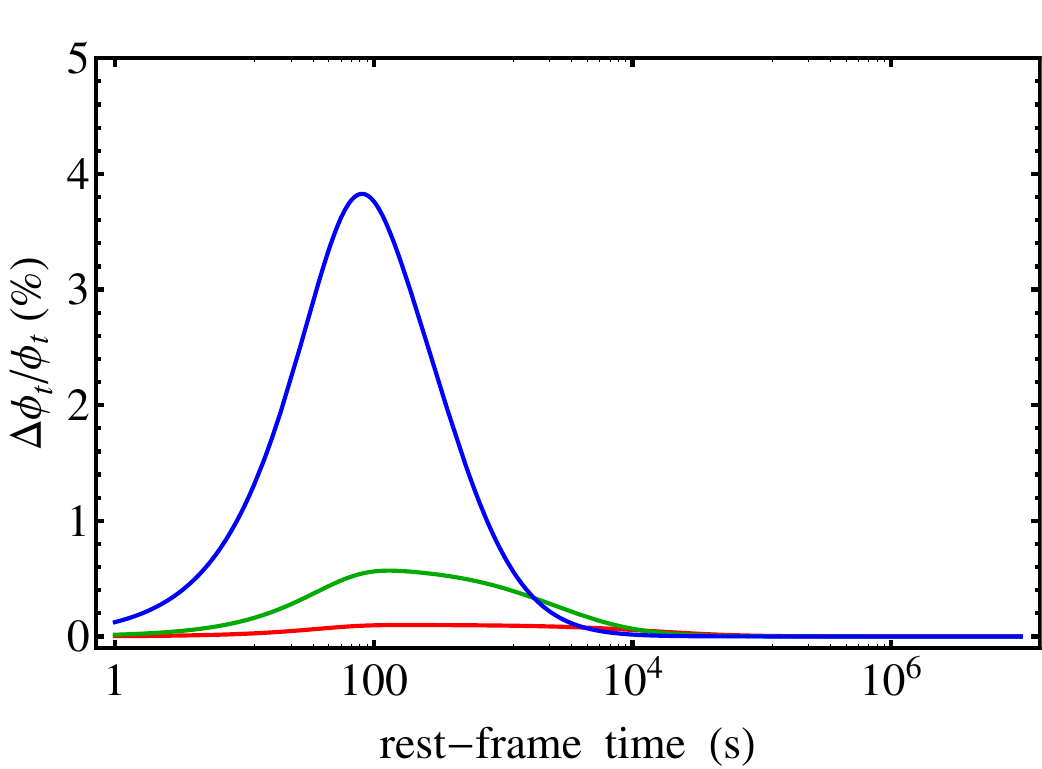}
\caption{Accuracies of the approximated solutions $\Delta \phi_n(t)/\phi_n(t)$ (left panel) and $\Delta \phi_t(t)/\phi_t(t)$ (right panel) obtained for the same values selected for Fig.~\ref{fig:1}, marked with the same choice of colors.}
\label{fig:A1}
\end{figure}

If we label the approximation in Eq.~\eqref{muccillingale3} as $\tilde\phi_n(t)$, the accuracy of the approximation can be computed as
\begin{equation}
\nonumber
\frac{\Delta \phi_n(t)}{\phi_n(t)}=\left|\frac{\phi_n(t)-\tilde\phi_n(t)}{\phi_n(t)}\right|\,.
\end{equation}
The result is shown in the left panel of Fig.~\ref{fig:A1}, with the same colors and values selected for Fig.~\ref{fig:1}. It is clear that the approximation work pretty well within an accuracy of $\approx4\%$, which is small, compared to the typical relative errors attached to the X-ray data.

For the thermal part, the analytic solution of Eq.~(\ref{partial1}) is
\begin{equation}
\label{SN_Ibc}
\phi_t(t) = e^{-z} + \frac{\tau_0 e^w \left[\bar\Gamma(1+\frac{\alpha}{2},w) - \bar\Gamma(1+\frac{\alpha}{2},w+z)\right]}{\tau_t(w+z)^{\alpha/2}}\,,
\end{equation}
where we have defined 
\begin{equation}
w(t) = -\frac{(t+\tau_h)^2}{2\tau_0\tau_h}\quad,\quad z(t) = \frac{t}{\tau_0}\left(1+\frac{t}{2\tau_h}\right)\,.
\end{equation}
We resort again the expansion in Eq.~\eqref{delta_gamma}, holding now for $|w|>|z|$ (or $\tau_h>2\tau_0$) and retain the lower order to find
\begin{equation}
\label{SN_Ibc_bis} 
\phi_t(t) = e^{-z} + \frac{\tau_0}{\tau_t}\left(1-e^{-z}\right)\left(1+\frac{t}{\tau_h}\right)^\alpha\,,
\end{equation}
that corresponds to Eq.~\eqref{timepartsol}.

Also in this case, we indicate the solution in Eq.~\eqref{timepartsol} with $\tilde\phi_t(t)$ and compute the accuracy
\begin{equation}
\nonumber
\frac{\Delta \phi_t(t)}{\phi_t(t)}=\left|\frac{\phi_t(t)-\tilde\phi_t(t)}{\phi_t(t)}\right|\,.
\end{equation}
The right panel of Fig.~\ref{fig:A1} shows that
the accuracy is within $\approx4\%$, again, smaller than the typical relative errors attached to any afterglow thermal components.




\end{appendix}

\end{document}